\documentclass[aps, prb, twocolumn, amsmath,amssymb,showpacs,superscriptaddress]{revtex4-1}

\usepackage{graphicx}
\usepackage{color}
\usepackage{amsthm}
\usepackage{amsmath}
\usepackage{amssymb}
\usepackage{setspace}
\usepackage{braket}
\usepackage{float}
\renewcommand{\(}{\left(}
\renewcommand{\)}{\right)}

\makeatletter
\@ifundefined{textcolor}{}
{
 \definecolor{BLACK}{gray}{0}
 \definecolor{WHITE}{gray}{1}
 \definecolor{RED}{rgb}{1,0,0}
 \definecolor{GREEN}{rgb}{0,1,0}
 \definecolor{BLUE}{rgb}{0,0,1}
 \definecolor{CYAN}{cmyk}{1,0,0,0}
 \definecolor{MAGENTA}{cmyk}{0,1,0,0}
 \definecolor{YELLOW}{cmyk}{0,0,1,0}
}

\renewcommand{\H}{\mathcal{H}}
\newcommand{\Pp}{\mathcal{P}}
\newcommand{\T}{\mathcal{T}}

\begin{document}

\title{Minimal models for topological Weyl semimetals}

\author{Timothy M. McCormick}
\email[]{mccormick.288@osu.edu}
\affiliation{Department of Physics and Center for Emergent Materials, The Ohio State University, Columbus, OH 43210, USA}

\author{Itamar Kimchi}
\email[]{itamark@mit.edu}
\affiliation{Department of Physics, Massachusetts Institute of Technology, Cambridge, MA 02139, USA}

\author{Nandini Trivedi}
\email[]{trivedi.15@osu.edu}
\affiliation{Department of Physics and Center for Emergent Materials, The Ohio State University, Columbus, OH 43210, USA}

\date{\today}

\begin{abstract}
Topological Weyl semimetals (TWS) can be classified as type-I TWS, in which the density of states vanishes at the Weyl nodes, and type-II TWS where an electron and a hole pocket meet with finite density of states at the nodal energy.  The dispersions of type-II Weyl nodes are tilted and break Lorentz invariance, allowing for physical properties distinct from those in a type-I TWS. 
We present minimal lattice models for both time-reversal-breaking and inversion-breaking type-II Weyl semimetals, and investigate their bulk properties and topological surface states. 
These lattice models capture the extended Fermi pockets and the connectivities of Fermi arcs. In addition to the Fermi arcs, which are topologically protected, we identify surface ``track states" that arise out of the topological Fermi arc states at the transition from type-I to type-II with multiple Weyl nodes, and persist in the type-II TWS. 
 \end{abstract}
\maketitle

%INTRODUCTION
\section{Introduction}

The band theory of solids was revolutionized by the discovery of topological insulators\cite{hasanKane,qiZhang}. The abundant list of topologically nontrivial quadratic Hamiltonians has been extended by the recent discovery of topological Weyl semimetals (TWS).  These materials have band crossings, at isolated points in momentum space, between two non-degenerate bands. The resulting nodes appear analogous to the Dirac nodes of graphene\cite{graphene}, but here exist in three dimensions rather than two. The three linearly independent momenta couple to all three Pauli matrices in the Hamiltonian, hence perturbations can shift the position of the node in momentum space but cannot open a gap.

%In 1929, Hermann Weyl first proposed the massless solution to the Dirac equation,\cite{Weyl1929} but there have been no known examples of Weyl fermions as fundamental particles.  However, 
There have been many recent theoretical proposals for the emergence of Weyl nodes in the band structure of solid state materials\cite{Bevan1997,Hsieh2008,
burkBal,wanTurnVish,PhysRevLett.107.186806,
Volovik2014514,1603.04744}.
In such a TWS, breaking either inversion or time reversal symmetry results in a pair of Dirac nodes separating into Weyl nodes.  These Weyl nodes are monopoles of Berry curvature in the Brillouin zone and the charge associated with such a monopole is known as its chirality.  Weyl nodes must come in pairs of opposite chirality\cite{Nielsen1981219} such that the net chirality over the Brillouin zone is zero.  A consequence of these bulk Weyl nodes is the existence of topological Fermi arcs on the surface of a TWS~\cite{wanTurnVish}.

Both the bulk Weyl nodes and the surface Fermi arcs have unique signatures in angle resolved photoemission spectroscopy (ARPES) experiments.  Searching for these signatures has proven to be extremely fruitful and several groups\cite{Xu613,PhysRevX.5.031013,Lv2015,Xu2015,Xue1501092,
Liu2016,PhysRevLett.116.066802} have discovered a TWS phase in the transition metal pnictide family: TaAs, TaP, NbP and NbAs.  These materials all belong to the so-called type-I TWS phase where the Weyl points are formed from a direct gap semiconductor closing linearly at a discrete set of Weyl points.  A separate class, known as type-II TWS, was recently predicted to arise from an indirect-gap semimetal, with the direct gap closing linearly at the Weyl nodes.  These predictions have been made for a variety of compounds\cite{Soluyanov2015,1511.07440,
1603.04624,Chang2016}.  Recently, signatures of a type-II TWS have been reported\cite{1512.09099,1603.06482,1603.07318} in  Mo$_x$W$_{1-x}$Te$_2$, stoichiometric MoTe$_2$, and LaAlGe, opening the door for further experimental study of the type-II TWS.

Although there have been some studies of lattice models for type-I TWS,\cite{0295-5075-97-6-67004,PhysRevB.87.245112,PhysRevB.91.115203} much of the theoretical work on topological Weyl semimetals has focused on low energy effective models of single Weyl nodes.  In a type-I TWS, where the density of states vanishes at the energy of the Weyl nodes, these effective models capture much of the essential physics including electro- and magnetotransport, \cite{vishChargexport,sonSpivakWeyl,PhysRevB.88.125105,PhysRevLett.111.246603,PhysRevLett.111.027201,
Kargarian2015,PhysRevB.93.085442,2014/10/20/online,PhysRevB.93.085107} thermoelectric properties, \cite{fieteThermoelec,PhysRevB.93.035116,PhysRevB.92.205113,PhysRevB.92.075205} magnetic properties,\cite{PhysRevB.93.045201} and effects of disorder\cite{PhysRevLett.114.257201,PhysRevB.92.174202,PhysRevB.93.075113}.  In a type-I TWS, when the chemical potential is shifted slightly away from the nodal energy, the Fermi pockets enclosing the projections of the Weyl nodes are very small. However, in a type-II TWS extended pockets of holes and electrons exist at the node energy.  It is clear then that small shifts to the chemical potential can result in the projections of the Weyl nodes being enclosed in comparatively large pockets in contrast to small pockets of the doped type-I case.  Fermi pockets at the node energy are among the defining characteristics of the type-II TWS, and a comprehensive study of the interplay of these bulk Fermi pockets, the nodes, and topological Fermi arcs requires a model that encompasses the entire Brillouin zone.  We present here the first thorough study of a variety of lattice models for type-II TWS.

Our paper is organized as follows.  In Sec. II, we outline the general form of the dispersion of a type-II Weyl node and summarize the symmetry properties that a TWS must obey.  In Sec. III, we consider the minimal ``Hydrogen atom" model for a time reversal breaking type-II TWS with a single pair of Weyl nodes and a single hole and electron pocket. We discuss its bulk band structure as well as possible configurations of Fermi arcs.  We show that in this minimal model the surface states are not topological.  We generalize to the ``Helium atom" model with isolated pairs of Fermi pockets around each Weyl node and show that this model does indeed contain topologically protected Fermi arcs.  In Sec. IV, we present an inversion breaking model for a type-II Weyl semimetal.  We find that this four node model has a rich set of possible Fermi arcs as well as supporting a surface state that we call the ``track state." Sec. V contains a discussion of the types of surface states supported by our lattice models and we conclude in Sec. VI with prospects for future investigations.

\section{General Considerations}

The defining features of a TWS are the nodal energy crossings in the Brillouin zone, so a minimal lattice model for a TWS must have at least two bands of the form
\begin{equation}
\hat{H} = \sum_{\mathbf{k}} \hat{c}^{\dagger}_{\mathbf{k}\alpha} \( \hat{\H}(\mathbf{k}) \)_{\alpha \beta} \hat{c}_{\mathbf{k}\beta}
\label{2nodeHam}
\end{equation}
where $\hat{c}^{(\dagger)}_{\mathbf{k}\alpha}$ annihilates (creates) an electron at momentum $\mathbf{k}$ in orbital $\alpha$ and
\begin{equation}
\label{gen2band}
\hat{\H}(\mathbf{k}) = \sum_{i=0,1,2,3} d_{i}(\mathbf{k})\ \hat{\sigma}_{i}.
\end{equation}
Here $\hat{\sigma}_{i}$ is the $i$-th Pauli matrix for $i = 1,2,3$ whose indices correspond to an orbital degree of freedom and $\hat{\sigma}_{0}$ is the $2\times2$ identity matrix.  If such a Hamiltonian has at least two points around which the Hamiltonian is described locally by 
\begin{equation}
\label{genWP}
\hat{\H}_{\textrm{WP}}(\mathbf{k}) = \sum_{i=1,2,3} \gamma_{i}k_{i}\hat{\sigma}_{0}+ \sum_{i,j=1,2,3} k_{i} A_{ij} \hat{\sigma}_{i},
\end{equation}
it describes a Weyl semimetal with nodes of chirality $\chi = \textrm{det}(A_{ij})$.
It is straightforward to show that the energy spectrum for the Hamiltonian in Eqn. (\ref{genWP}) is given by 
\begin{multline}
\label{genWPnrg}
E_{\pm}(\mathbf{k}) = \sum_{i=1,2,3} \gamma_{i}k_{i} \pm \sqrt{\sum_{j=1,2,3} \left( \sum_{i=1,2,3}
k_{i} A_{ij}
\right)^2} \\ = T(\mathbf{k}) \pm U(\mathbf{k}),
\end{multline}
where $T(\mathbf{k})$ tilts the Weyl cone.  The definition\citep{Soluyanov2015} of a type-II Weyl node is one where there exists a direction $\mathbf{e}_{\mathbf{k}}$ in the Brillouin zone such that
\begin{equation}
 T(\mathbf{e}_{\mathbf{k}}) > U(\mathbf{e}_{\mathbf{k}}).
 \label{typeIIcond}
\end{equation}

Since in the presence of both inversion and time reversal symmetry the Berry curvature is identically zero throughout the Brillouin zone, the presence of Weyl nodes relies on breaking either inversion (henceforth labeled $\hat{\Pp}$) or time reversal symmetry (labeled $\hat{\T}$).  For spinless fermions, we choose a definite representation for the $\hat{\Pp}$ and $\hat{\T}$ operators,
\begin{equation}
\label{symmOpsSpinless}
\hat{\Pp} \leftrightarrow \hat{\sigma}_1,\ \hat{\T} \leftrightarrow \hat{K},
\end{equation}
where $\hat{K}$ is the anti-Hermitian complex conjugation operator.  Each of $\hat{\Pp}$ and $\hat{\T}$ also reverse the sign of the momentum such that $\mathbf{k} \rightarrow -\mathbf{k}$.  In this paper we investigate lattice models for Weyl semimetals that break either $\hat{\T}$ or $\hat{\Pp}$, and using the definitions in Eqn. (\ref{symmOpsSpinless}) it will be straightforward to show this symmetry breaking explicitly for each model we consider. 

\section{Time Reversal Breaking Model}

We begin by investigating a lattice model given by a Hamiltonian $\hat{\H}(\mathbf{k})$ that hosts Weyl nodes and breaks time reversal symmetry but preserves inversion symmetry such that
\begin{equation}
\label{trbSymcond}
\hat{\Pp}^{\dagger}\hat{\H}(-\mathbf{k}) \hat{\Pp} =  \hat{\H}(\mathbf{k}),\ \hat{\T}^{\dagger}\hat{\H}(-\mathbf{k}) \hat{\T} \neq  \hat{\H}(\mathbf{k}).
\end{equation}
The minimal number of Weyl nodes for such a Hamiltonian is two and we find that such a minimal model can be used to investigate a wide range of possible TWS Fermi surface and arc connectivity. We begin by writing down the simplest possible two node time-reversal breaking (TRB) Hamiltonian with a type-II tilt and investigating its band structure.  A pair of Weyl nodes are formed from the nodal crossing of exactly one electron band with one hole band.  By calculating the band structure for the system in a finite slab geometry, we investigate the surface Fermi arc behavior.  We then show that this minimal model can be modified with a term that splits these electron and hole pockets into pairs that exist around each node.

\subsection{The ``Hydrogen atom" for a type II time reversal breaking TWS}

\begin{figure*}
	\centering
	\includegraphics[width=1.\textwidth]{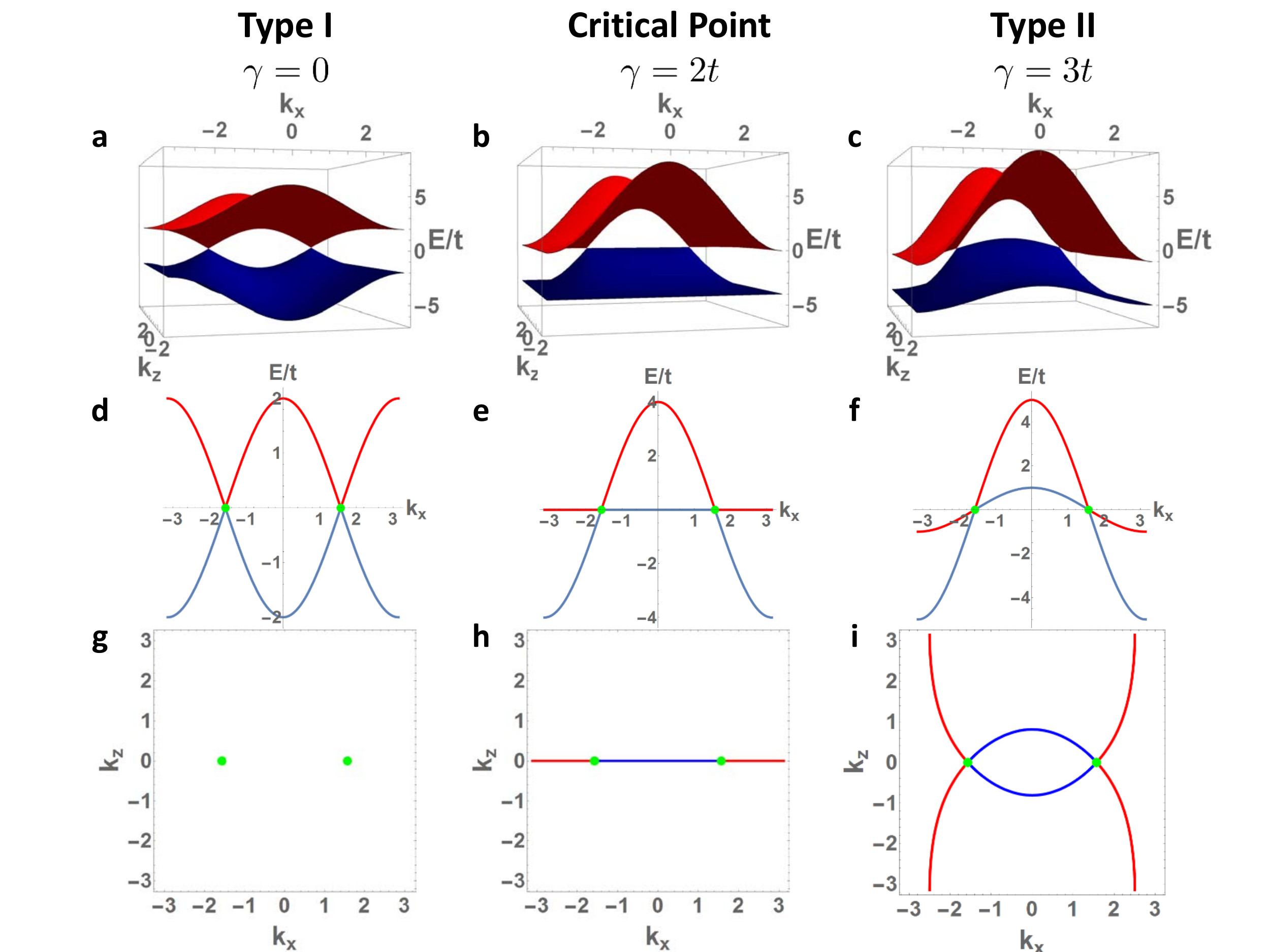}
	\caption{{\bf{Bulk band structure for the ``Hydrogen atom" of type-I and type-II Weyl semimetal. }}
{\bf{a-c}} The bulk band structure for the Hamiltonian in Eqn. (\ref{type2simpleTRB}) at $k_y = 0$ with parameters $k_0 = \pi/2$, $t_{x} 	 = t$, $m = 2t$ for ({\bf{a}}) type-I Weyl semimetal with $\gamma = 0$, ({\bf{b}}) the critical point between type-I and type-II Weyl semimetal with $\gamma = 2 t$ and ({\bf{c}}) type-II Weyl semimetal with $\gamma = 3 t$.
{\bf{d-f}} Cuts through the Weyl nodes at $k_y = k_z = 0$ for the same parameters as ({\bf{a-c}}). The cones comprising the Weyl nodes develop a characteristic tilt of the type-II TWS as $\gamma$ is increased.
{\bf{g-i}} Constant energy cuts through the nodal energy ($E = 0$) for the same parameters as ({\bf{a-c}}).  We see that for a type-I TWS, there are no states at the Fermi energy.  At the critical point between a type-I and type-II TWS, we see lines of bulk states appearing between the nodes.  These lines open into bulk pockets when the system becomes a type-II TWS.
	}
	\label{typeIvsII_TRB_bulkFig}
\end{figure*}

The following Hamiltonian 
\begin{multline}
\label{type2simpleTRB}
\hat{\H}^{\textrm{TRB}}_A(\mathbf{k}) =\gamma \big(\textrm{cos}(k_x)-\textrm{cos}(k_0)\big)\hat{\sigma}_{0}\\- \big(m(2-\textrm{cos}(k_y)-\textrm{cos}(k_z))+2t_x(\textrm{cos}(k_x)-\textrm{cos}(k_0))
\big)\hat{\sigma}_1\\
-2t\ \textrm{sin}(k_y) \hat{\sigma}_2-2t\ \textrm{sin}(k_z) \hat{\sigma}_3
\end{multline}
satisfies the symmetry conditions in Eqn. (\ref{trbSymcond}) and possesses two Weyl nodes at $\mathbf{k} = (\pm k_0,0,0)$.  When $\gamma = 0$, this Hamiltonian is known\cite{luRanWSM} to host nodes of type-I.  However, the addition of the term $\gamma \big(\textrm{cos}(k_x)-\textrm{cos}(k_0)\big)\hat{\sigma}_{0}$ bends both bands and when $\gamma > 2t_x$ it is simple to see these nodes become	 type-II as defined by Eqn. (\ref{typeIIcond}).  We see this evolution from type-I to type-II very clearly in Fig. \ref{typeIvsII_TRB_bulkFig}.  When $\gamma = 0$, the hole band (blue) touches the electron band (red) at the two Weyl points where the density of states vanishes, as seen in Fig. \ref{typeIvsII_TRB_bulkFig}a,d,g.  When the system is in the type-II regime, the Weyl cones are tilted and this leads to a nonzero density of electron and hole states at the node energy, as seen clearly in Fig. \ref{typeIvsII_TRB_bulkFig}c,f,i.  When $\gamma = 2t_x$ exactly, the system is at a critical point between a type-I and a type-II Weyl semimetal.  This is clearly seen in Fig. \ref{typeIvsII_TRB_bulkFig}b,e,h, where a single line of bulk states connect the Weyl points at $E = 0$.  The states seen in Fig. \ref{typeIvsII_TRB_bulkFig}h open up into the electron and hole pockets seen at $E = 0$ for the type II case in Fig. \ref{typeIvsII_TRB_bulkFig}i.

In a type-II TWS, it is important to consider the net chirality enclosed by the bulk Fermi pockets when determining the Fermi arc connectivity.  If one encloses a bulk pocket by a Gaussian surface in a region where the band structure is gapped, the number of Fermi arcs impinging on the Gaussian surface are quantized and equal to the net chirality of Weyl nodes enclosed. 
When the model in Eqn. (\ref{type2simpleTRB}) is in the type-II regime and the chemical potential is shifted away from $E = 0$, the projections of both Weyl nodes are either enclosed in the electron pocket ($E > 0$) or the projections are both enclosed in the hole pocket ($\mu < 0$).  Since the projections of both nodes lie within the same Fermi pocket, we expect that Fermi arcs in this system are not topologically protected in general.  Surface states may exist, but their lack of topological protection stems from the fact that there are no isolated Fermi pockets that enclose Weyl nodes with nonzero net chirality.  As a result, the surface states can hybridize with bulk states and are therefore trivial.

\begin{figure*}
	\centering
	\includegraphics[width=1.\textwidth]{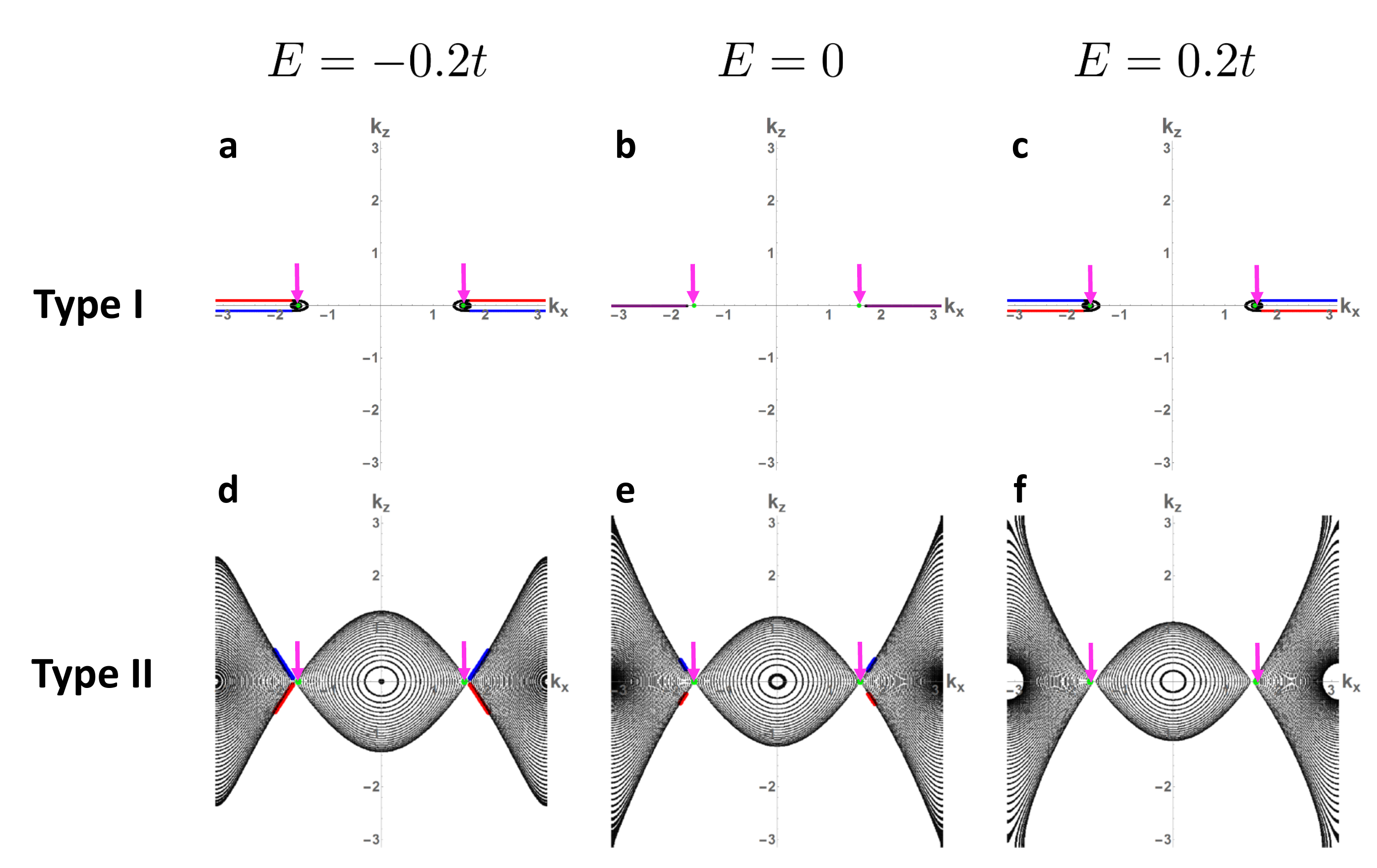}
	\caption{{\bf{Fermi surface and arc configuration for the ``Hydrogen atom" of type-I and type-II TWS. }}
{\bf{a-c}} Bulk Fermi surfaces and surface Fermi arcs for a type I TWS with the same bulk parameters as in Fig. \ref{typeIvsII_TRB_bulkFig}a,d,g calculated in a slab geometry with $L = 50$ layers in the $y$-direction.  The slab calculations are done at the following constant energy: {\bf{(a)}} $E = -0.2t$, {\bf{(b)}} $E = 0$, {\bf{(c)}} $E = 0.2t$.  We color the states exponentially localized to the $y = 1$ ($y = L$) surface red (blue) and note that such surface states form topological arcs connecting the two Weyl nodes (shown as green dots and marked with pink arrows).  We note that at $E = 0$ the two Fermi arcs are degenerate along $k_z = 0$ and we color them purple to signify this.
{\bf{d-f}} Bulk Fermi surfaces and surface Fermi arcs for a type-II TWS with the same bulk parameters as in Fig. \ref{typeIvsII_TRB_bulkFig}c,f,i calculated in a slab geometry with $L = 50$ layers in the $y$-direction.  The slab calculations are done at the same constant energies as above: {\bf{(d)}} $E = -0.2t$, {\bf{(e)}} $E = 0$, {\bf{(f)}} $E = 0.2t$.
	}
	\label{simpTRB_FS}
\end{figure*}

In order to investigate the structure of the Fermi arcs, we introduce an edge by considering a slab with a finite thickness in one direction.  We partially Fourier transform the Hamiltonian in Eqn. (\ref{type2simpleTRB}) into real space for a $L$ layer system in the $y$-direction, while keeping the system infinite in the $x$- and $z$-directions.  In Fig. \ref{simpTRB_FS}, we show the results of such a slab calculation for the model given by Eqn. (\ref{type2simpleTRB}) in the type I regime ($\gamma = 0$) with the same bulk parameters as in Fig. \ref{typeIvsII_TRB_bulkFig}a,d,g and in the type II regime ($\gamma = 3t_x$) with the same bulk parameters as in Fig. \ref{typeIvsII_TRB_bulkFig}c,f,i for $L = 50$ layers.  We calculate the expectation of the finite position $\braket{y}$ and label the states as ``top" (``bottom") if they are exponentially localized at $\braket{y} = 1$ ($\braket{y} = L$).  We color these top and bottom states red and blue respectively.

As we expect, for the type-I case when $\gamma = 0$, a Fermi arc on each surface connects the Weyl nodes, as seen in Fig. \ref{simpTRB_FS}a-c.  This is seen clearly in Fig. \ref{simpTRB_FS}b where two Fermi arcs connect the two nodes from $(k_x,k_z) = (-\pi/2,0)$ to $(k_x,k_z) = (\pi/2,0)$.  At $E = 0$, both the top and bottom arcs are degenerate at $k_z = 0$, shown as a purple line.  When we lower the Fermi energy below the node energy, each node is enclosed in a small isolated Fermi pocket.  Since each pocket encloses a net chirality $\chi = \pm 1$, the pockets are connected by an arc on each surface, as seen in Fig. \ref{simpTRB_FS}a.  The same is seen at higher energies $E > 0$ in Fig. \ref{simpTRB_FS}c.

We calculate the band structure in the slab geometry for a type-II TWS ($\gamma = 3t_x$) and find that there are marked differences in the surface state behavior (see Fig. \ref{simpTRB_FS}d-f).  Since both nodes are formed from  a single electron and a single hole pocket, we cannot construct a simply connected 2D Gaussian surface in the Brillouin zone that encloses a single node.  When the energy is lower than the Weyl energy in Fig. \ref{simpTRB_FS}d, we see that the projections of both nodes are enclosed by the same hole pocket. Although there are two sets of surface states connecting the hole and electron pockets, they are trivial in a topological sense. When one considers a Gaussian surface that encloses the central hole pocket, it is pierced by four arcs, two on each real-space surface.  The Fermi velocity of each arc is opposite on a given real-space surface and so the net chirality of the arcs is zero.  We see that as we raise the chemical potential to the node energy and above, these arcs disappear completely.  This is completely different from the type-I case where the arcs exist at all energies since the nodes were always isolated in separate Fermi pockets.

\subsection{The ``Helium atom" for a type II time reversal breaking TWS}

\begin{figure*}
	\centering
	\includegraphics[width=1.\textwidth]{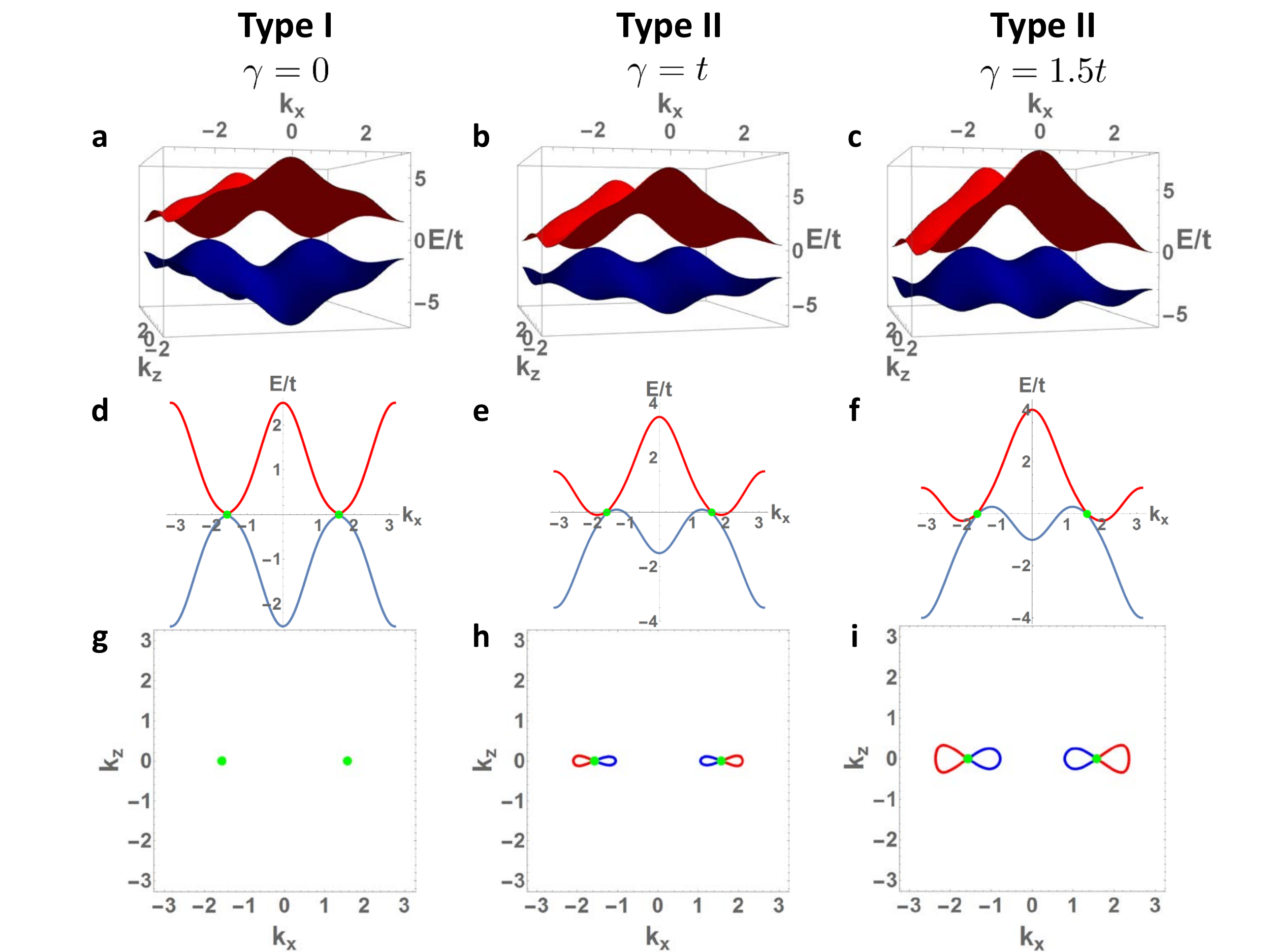}
	\caption{{\bf{Bulk band structure for type-I and type-II TRB model with separate pockets (the ``Helium atom"). }}
{\bf{a-c}} The bulk band structure for the Hamiltonian in Eqn. (\ref{2nodeSepKernel}) at $k_y = 0$ with the parameters $k_0 = \pi/2$, $t_{x} = t$, $m = 2t$ and $\gamma_x = t/2$ for ({\bf{a}}) type-I TWS with $\gamma = 0$, ({\bf{b}}) type-II TWS with $\gamma = t$ and ({\bf{c}}) type-II TWS with $\gamma = 1.5 t$.
{\bf{d-f}} Cuts through the Weyl nodes at $k_y = k_z = 0$ for the same parameters as ({\bf{a-c}}). The cones comprising the Weyl nodes again develop a characteristic tilt of the type-II TWS as $\gamma$ is increased.  
{\bf{g-i}} Constant energy cuts through the nodal energy ($E = 0$) for the same parameters as ({\bf{a-c}}).  Note that for a type-I TWS, there are no states at the Fermi energy while in the
type-II regime, there are two sets of electron and hole pockets on either side of the Weyl nodes.
	}	\label{typeIvsII_sepPockets_TRB_bulkFig}
\end{figure*}

It is straightforward to add a term to the Hamiltonian in Eqn. (\ref{type2simpleTRB}) that results in a lattice model for a type-II TWS with hole and electron pockets that do not enclose both Weyl nodes.  Due to the pairs of electron and hole pockets supported by this model, we call it the ``Helium model" for a type-II time-reversal-breaking TWS in analogy with the ``Hydrogen model" above. We consider the following Hamiltonian 
\begin{equation}
\hat{\H}^{\textrm{TRB}}_B(\mathbf{k}) = \hat{\H}^{\textrm{TRB}}_A(\mathbf{k}) -\gamma_x (\textrm{cos}(3 k_x)-\textrm{cos}(3 k_0))\hat{\sigma}_1,
\label{2nodeSepKernel}
\end{equation}
where we have added to Eqn. (\ref{type2simpleTRB}) the term proportional to $\gamma_x$.  In general, this model supports up to six Weyl nodes.  However, so long as $|2 t_x| > |3 \gamma_x|$, there are only two Weyl nodes in the Brillouin zone.  These nodes are located at $E = 0$ and $\mathbf{k} = (\pm k_0,0,0)$ and they are type-II if $\gamma > 3 \gamma_x -2 t_x$.  The addition of the term $\gamma_x (\textrm{cos}(3 k_x)-\textrm{cos}(3 k_0))$ gaps out the bulk spectrum along the lines $(k_y,k_z) = (0,0)$ and $(k_y,k_z) = (0,\pi)$ at the nodal energy.  This leads to a pair of isolated hole pockets touching a pair of isolated electron pockets at the Weyl nodes when the system is type-II.  In Fig. \ref{typeIvsII_sepPockets_TRB_bulkFig}, we find that as $\gamma$ grows relative to $3 \gamma_x - 2 t_x$, the Fermi pockets grow in size.  This is because as the tilt of the nodes gets larger, more electron and hole states exist at the Fermi energy.  As we shift the chemical potential away from the node energy, the projections of the nodes are now isolated with each node in a single electron (hole) pocket when the chemical potential is raised (lowered).  

\begin{figure*}
	\centering
	\includegraphics[width=1.\textwidth]{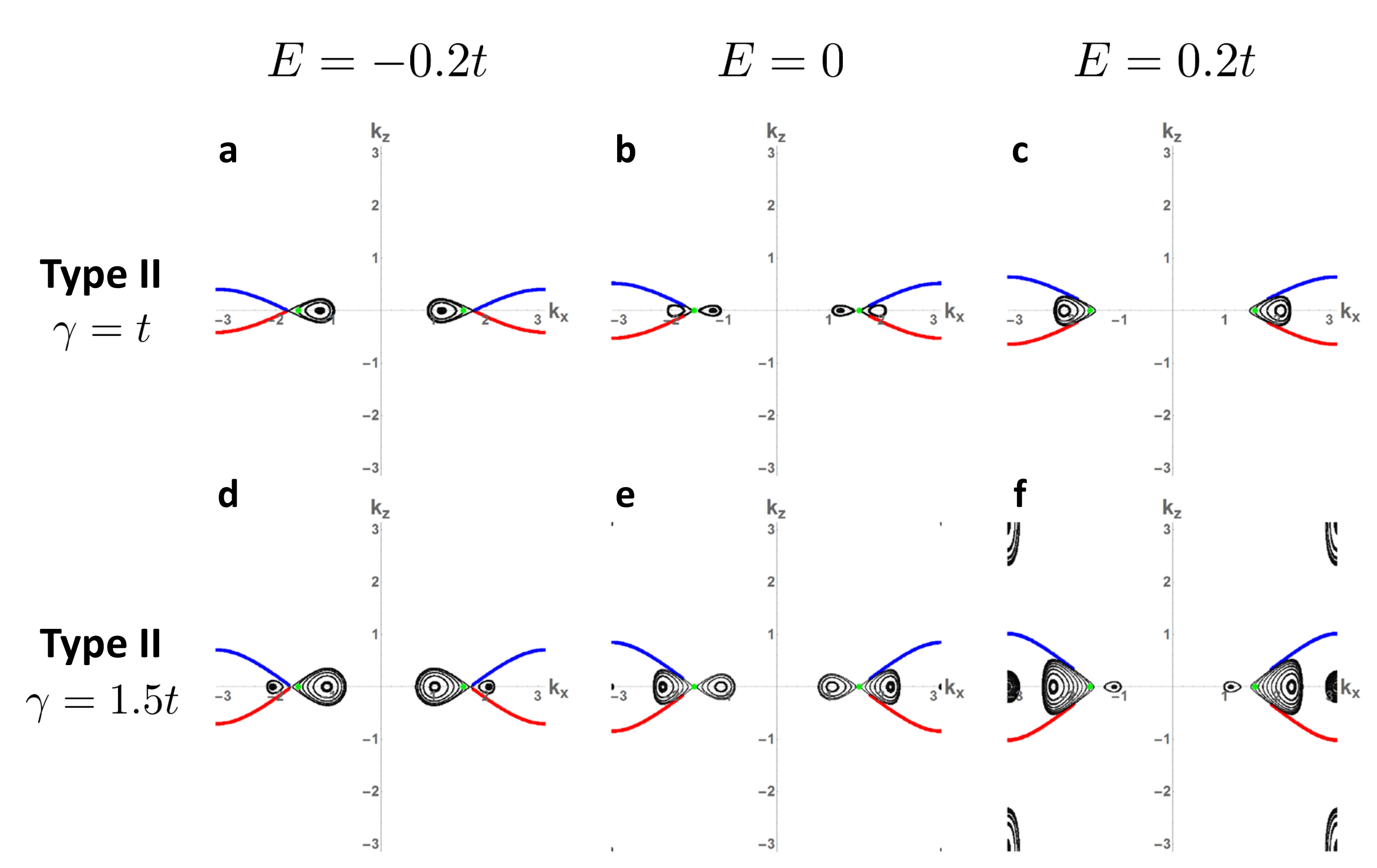}
	\caption{{\bf{Fermi surface and Fermi arc configuration for type I and type-II time-reversal-breaking model with separate pockets (the ``Helium atom"). }}
{\bf{a-c}} Bulk Fermi surfaces and surface Fermi arcs for a type-II Weyl semimetal given by Eqn. (\ref{2nodeSepKernel}) with the same bulk parameters as in Fig. \ref{typeIvsII_sepPockets_TRB_bulkFig}b,e,h calculated in a slab geometry with $L = 50$ layers in the $y$-direction.  The slab calculations are done at the constant energies: {\bf{(a)}} $E = -0.2t$, {\bf{(b)}} $E = 0$, {\bf{(c)}} $E = 0.2t$.  As in Fig. \ref{simpTRB_FS}, we color the states that are exponentially localized to the $y = 1$ ($y = L$) surface red (blue) and note that such surface states form topological arcs connecting the two Weyl nodes (shown as green dots).  We note unlike in Fig. \ref{simpTRB_FS}, each node is isolated in its own hole ({\bf{a}}) or electron ({\bf{c}}) pocket when the chemical potential is away from $E = 0$.  These pockets are connected by arcs confined to the surface in the $y$-direction. However, in this type-II TWS the Fermi pockets enclosing a Weyl node can be quite extended, unlike a type-I TWS, the arcs can terminate on a pocket quite far away from the projection of the nodes.
{\bf{d-f}} Bulk Fermi surfaces and surface Fermi arcs for a type-II TWS with the same bulk parameters as in Fig. \ref{typeIvsII_sepPockets_TRB_bulkFig}c,f,i calculated in a slab geometry with $L = 50$ layers in the $y$-direction.  The slab calculations are done at the same constant energies as above: {\bf{(d)}} $E = -0.2t$, {\bf{(e)}} $E = 0$, {\bf{(f)}} $E = 0.2t$.  We see that as the tilt grows, so do the pockets enclosing the nodes.  We note that a trivial electron pocket appears around the $(k_x,k_z) = (\pi,\pi)$ point.  This pocket encloses no Weyl nodes and so is not connected via Fermi arcs to any other pockets.
	}
	\label{sepTRB_FS}
\end{figure*}

We again consider the slab geometry described in the section above in order to investigate the interplay of the bulk pockets and the Fermi arcs for the model given by Eqn. (\ref{2nodeSepKernel}).  Unlike the simpler model described by Eqn. (\ref{type2simpleTRB}), we see in Fig. \ref{sepTRB_FS} that Eqn. (\ref{2nodeSepKernel}) supports isolated Fermi pockets enclosing the Weyl nodes in the type-II regime when $\gamma = t$ (Fig. \ref{sepTRB_FS}a-c) and $\gamma = 1.5 t$ (Fig. \ref{sepTRB_FS}d-f).  Unlike the Fermi surfaces and arcs generated by Eqn. (\ref{type2simpleTRB}), in Fig. \ref{sepTRB_FS} we see that each node is isolated in its own hole (Fig. \ref{sepTRB_FS}a,d) or electron (Fig. \ref{sepTRB_FS}c,f) pocket when the chemical potential is away from $E = 0$.  We emphasize that this is due to the extra $\hat{\sigma}_1$ term in the Hamiltonian in Eqn. (\ref{2nodeSepKernel}).  These isolated pockets in Fig. \ref{sepTRB_FS} are connected by arcs confined to the surface in the $y$-direction. However, in this type-II TWS the Fermi pockets enclosing a Weyl node can be quite extended and, unlike a type-I TWS, the arcs can terminate on a pocket quite far away from the projection of the nodes.  We see that as the tilt grows in Fig. \ref{sepTRB_FS}d-f, so do the pockets enclosing the nodes.  We note that a trivial electron pocket appears around the $(k_x,k_z) = (\pi,\pi)$ point.  This pocket encloses no Weyl nodes and therefore it is not connected via Fermi arcs to any other pockets.

Although the local linearized Hamiltonian describing the spectrum close to a node in Eqn. (\ref{2nodeSepKernel}) is identical to the effective Hamiltonian of nodes of the model described by Eqn. (\ref{type2simpleTRB}), the full lattice models describe topologically distinct configurations of bulk Fermi surfaces and surface Fermi arcs.  When there is only one electron pocket and one hole pocket with the projections of the Weyl nodes enclosed by the same pocket, the topological protection of the Fermi arcs is lost.  However, we see that once each node is enclosed in its own isolated pocket, the topological protection of the Fermi arcs is restored.

\begin{figure*}
	\centering
	\includegraphics[width=1.\textwidth]{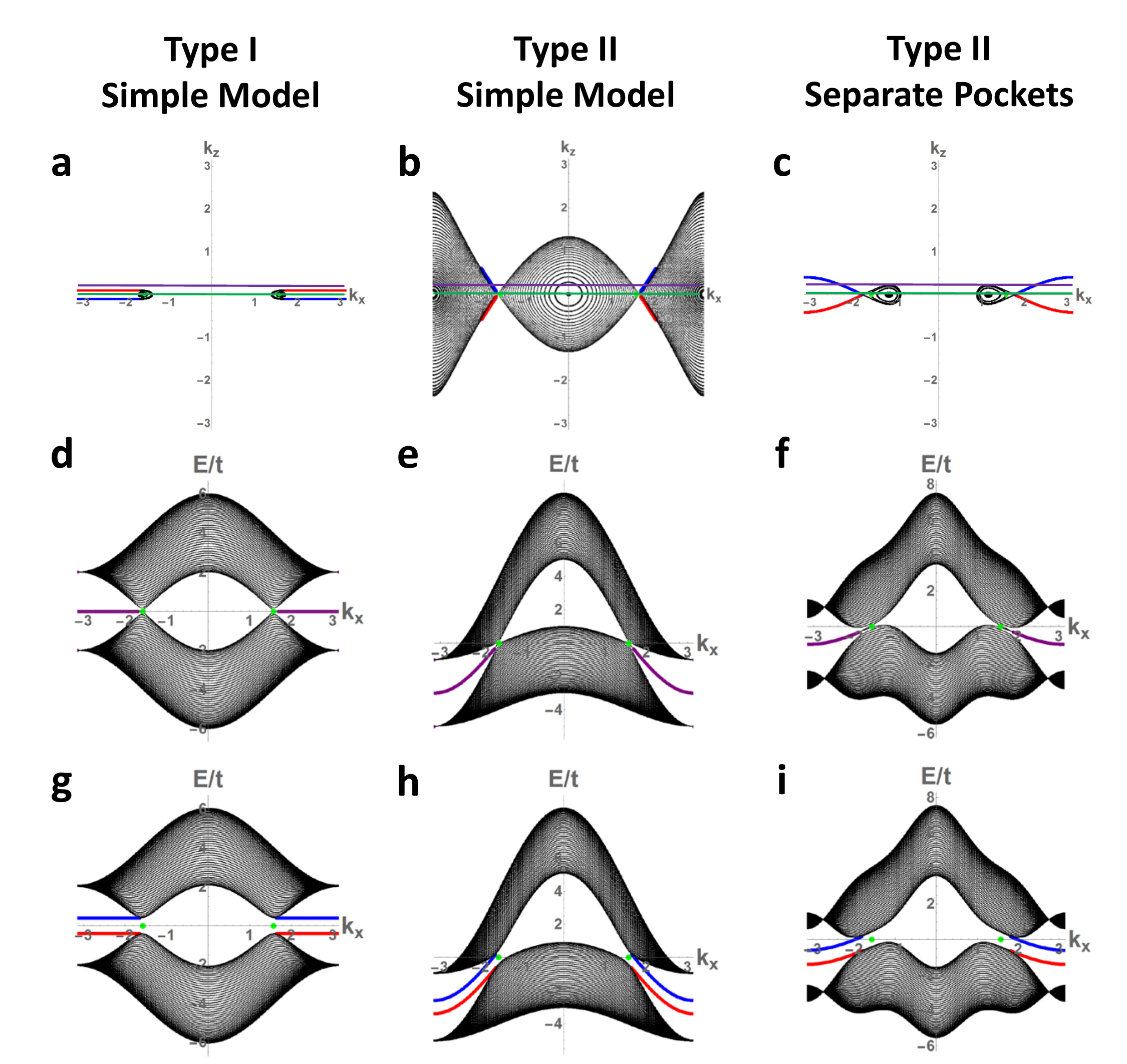}
	\caption{{\bf{Energy dispersion for type-I and type-II time-reversal-breaking Weyl semimetal showing Fermi arcs. }}
{\bf{a-c}} Fermi surface and Fermi arc configuration at $E = -0.2t$ for the simplest type-I case given by Eqn. (\ref{type2simpleTRB}) with $\gamma = 0$ ({\bf{a}}), the simple two pocket type-II case with $\gamma = 3t$ ({\bf{b}}), and the two node TRB case with isolated pockets surrounding each node with $\gamma = t$.  The green lines correspond to the cuts at $k_z = 0$ given in  ({\bf{d-f}}) and the purple lines correspond to the cuts at $k_z = 0$ given in  ({\bf{g-i}}).
{\bf{d-f}} Energy as a function of $k_x$ for slab calculation for the green cut above.  Top and bottom states are degenerate and shown in purple.
{\bf{g-i}} Energy as a function of $k_x$ for slab calculation for the purple cut above.  Top and bottom states are nondegenerate and shown as red and blue respectively.
	}
	\label{trbCuts}
\end{figure*}

Finally, we consider the energy dispersion of the Fermi arcs near a node.  Again using the slab geometry as above, we calculate the energy spectrum, this time at a constant $k_z$, as shown in Fig.  \ref{trbCuts}.  We see that for the simplest type-I case (Eqn. (\ref{type2simpleTRB}) with $\gamma = 0$), the surface arcs do not disperse in $k_x$ for a fixed $k_z$.  This changes in the type-II case for both the simple Hamiltonian in Eqns. (\ref{type2simpleTRB}) and (\ref{2nodeSepKernel}).  At fixed $k_z$, the arcs connecting the node inherit the tilt proportional to $\gamma$ and now bend.  This characteristic bend of the Fermi arc dispersion has been observed in ARPES studies of type-II Weyl semimetal\cite{1603.06482}.

\section{Inversion Breaking Model}

We now turn to a lattice model for a topological Weyl semimetal that breaks inversion symmetry but is invariant under time-reversal.  Analogous with Eqn. (\ref{trbSymcond}), we seek a Hamiltonian $\hat{\H}(\mathbf{k})$ that satisfies the following symmetry conditions
\begin{equation}
\label{invSymCond}
\hat{\Pp}^{\dagger}\hat{\H}(-\mathbf{k}) \hat{\Pp} \neq \hat{\H}(\mathbf{k}),\ \hat{\T}^{\dagger}\hat{\H}(-\mathbf{k}) \hat{\T} =  \hat{\H}(\mathbf{k}),
\end{equation}
where $\hat{\Pp}$ and $\hat{\T}$ are again given by Eqn. (\ref{symmOpsSpinless}).
Unlike a time-reversal-breaking Weyl semimetal, the minimum number of Weyl nodes for a spinless inversion-breaking (IB) TWS is four. 
More importantly, the lattice model for an IB TWS exhibits what we term "track states" that are loops of states that live on the surface of the TWS and are degenerate with the states forming the topological Fermi arcs.  However, unlike topological Fermi arcs, these track states form closed contours rather than open ones; they are not topological, but do evolve from the topological arc states upon the transition from type-I to type-II. 

It is easy to show that the Hamiltonian
\begin{multline}
\hat{\H}^{\textrm{IB}}(\mathbf{k}) =\gamma (\textrm{cos}(2 k_x)-\textrm{cos}(k_0))(\textrm{cos}(k_z)-\textrm{cos}(k_0))\hat{\sigma}_{0}\\- (m(1-\textrm{cos}^{2}(k_z)-\textrm{cos}(k_y))+2t_x(\textrm{cos}(k_x)-\textrm{cos}(k_0)))\hat{\sigma}_1
\\-2t\ \textrm{sin}(k_y) \hat{\sigma}_2-2t\ \textrm{cos}(k_z) \hat{\sigma}_3
\label{hinvkernel}
\end{multline}
satisfies the conditions in Eqn. (\ref{invSymCond}).
When $\gamma = 0$, Eqn. (\ref{hinvkernel}) describes a TWS with four nodes located at $\mathbf{k}_{\textrm{W}} = (\pm k_0,0,\pm \pi/2)$ that breaks inversion but preserves time-reversal symmetry.  The term $\gamma (\textrm{cos}(2 k_x)-\textrm{cos}(k_0))(\textrm{cos}(k_z)-\textrm{cos}(k_0))\hat{\sigma}_{0}$ causes a different shift in both band than those considered in the time reversal breaking cases and results in both bands bending in both the $k_x$- and $k_z$-directions.  This can produce isolated Fermi pockets around the Weyl points without having to add an additional $\hat{\sigma}_1$ term like in the time-reversal-breaking case in Eqn. (\ref{2nodeSepKernel}).  The inversion-breaking model above also easily generates trivial Fermi pockets that exist in isolation from those that meet at the Weyl nodes.

\begin{figure*}
	\centering	\includegraphics[width=0.9\textwidth]{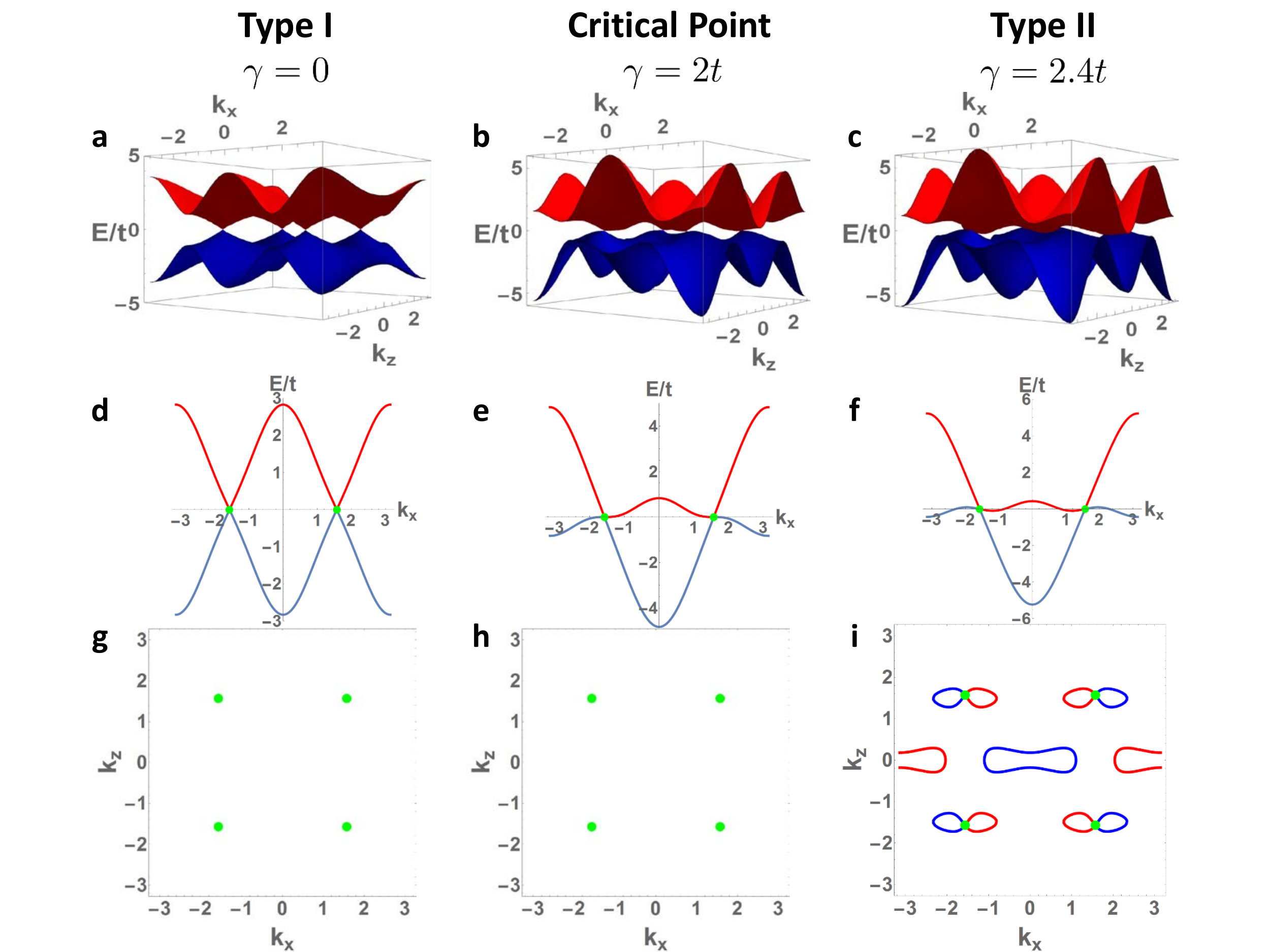}
	\caption{{\bf{Bulk band structure for type-I and type-II inversion breaking TWS. }}
	{\bf{a-c}} The bulk band structure for the Hamiltonian in Eqn. (\ref{hinvkernel}) at $k_y = 0$ with the parameters $k_0 = \pi/2$, $t_{x} = t/2$, $m = 2t$ for ({\bf{a}}) type I TWS with $\gamma = 0$, ({\bf{b}}) the critical point between a type-I and a type-II TWS with $\gamma = 2 t$ and ({\bf{c}}) type-II TWS with $\gamma = 2.4 t$.
	{\bf{d-f}} Cuts through the Weyl nodes at $k_y = 0$ and $k_z = -\pi/2$ for the same parameters as ({\bf{a-c}}).
	These cuts are shown as the green lines in ({\bf{g-i}}). 
	 The cones comprising the Weyl nodes develop a characteristic tilt of the type-II Weyl node as $\gamma$ is increased.
	 {\bf{g-i}} Constant energy cuts through the nodal energy ($E = 0$) for the same parameters as ({\bf{a-c}}).  We see that for a type-I Weyl semimetal, there are no states at the Fermi energy.  At the critical point between a type-I and type-II TWS, the density of states still vanishes.  In the type-II regime, electron and hole pockets form near the Weyl nodes.  These pockets enclose the projections of the Weyl nodes when the chemical potential is shifted away from $E = 0$.  Trivial pockets also appear at $\mathbf{k} = (0,0,0)$ and $\mathbf{k} = (0,0,\pi)$.
	}
	\label{invBulk}
\end{figure*}

We show the bulk band structure for Eqn. (\ref{hinvkernel}) in Fig. \ref{invBulk}.  We see that indeed when $\gamma = 0$ (Fig. \ref{invBulk}a,d,g), the electron band meets the hole band at four isolated type-I Weyl points and the density of states vanishes at the nodal energy.  As $\gamma$ increases, the Weyl nodes begin to tilt in the $k_z$-direction.  When $\gamma$ is tuned to the critical point between the type-I and type-II phases (Fig. \ref{invBulk}b,e,h), the electron and hole pockets still meet at the four Weyl nodes with a vanishing density of states, but we see in Fig. \ref{invBulk}e that the Weyl nodes are now tilting in the $k_z$-direction.  As $\gamma$ is further increased into the type-II limit (Fig. \ref{invBulk}c,f,i), we now see that the nodes are tilted as seen in Fig. \ref{invBulk}f and the electron (hole) pockets are shifted below (above) the node energy. In particular, we see in Fig. \ref{invBulk}i that there are four electron and four hole pockets that exist at $E = 0$ and meet at the Weyl nodes.  There is also a trivial hole pocket centered at $\mathbf{k} = (0,0,0)$ and a trivial electron pocket centered at $\mathbf{k} = (\pi,0,0)$.

\begin{figure*}
	\centering	\includegraphics[width=\textwidth]{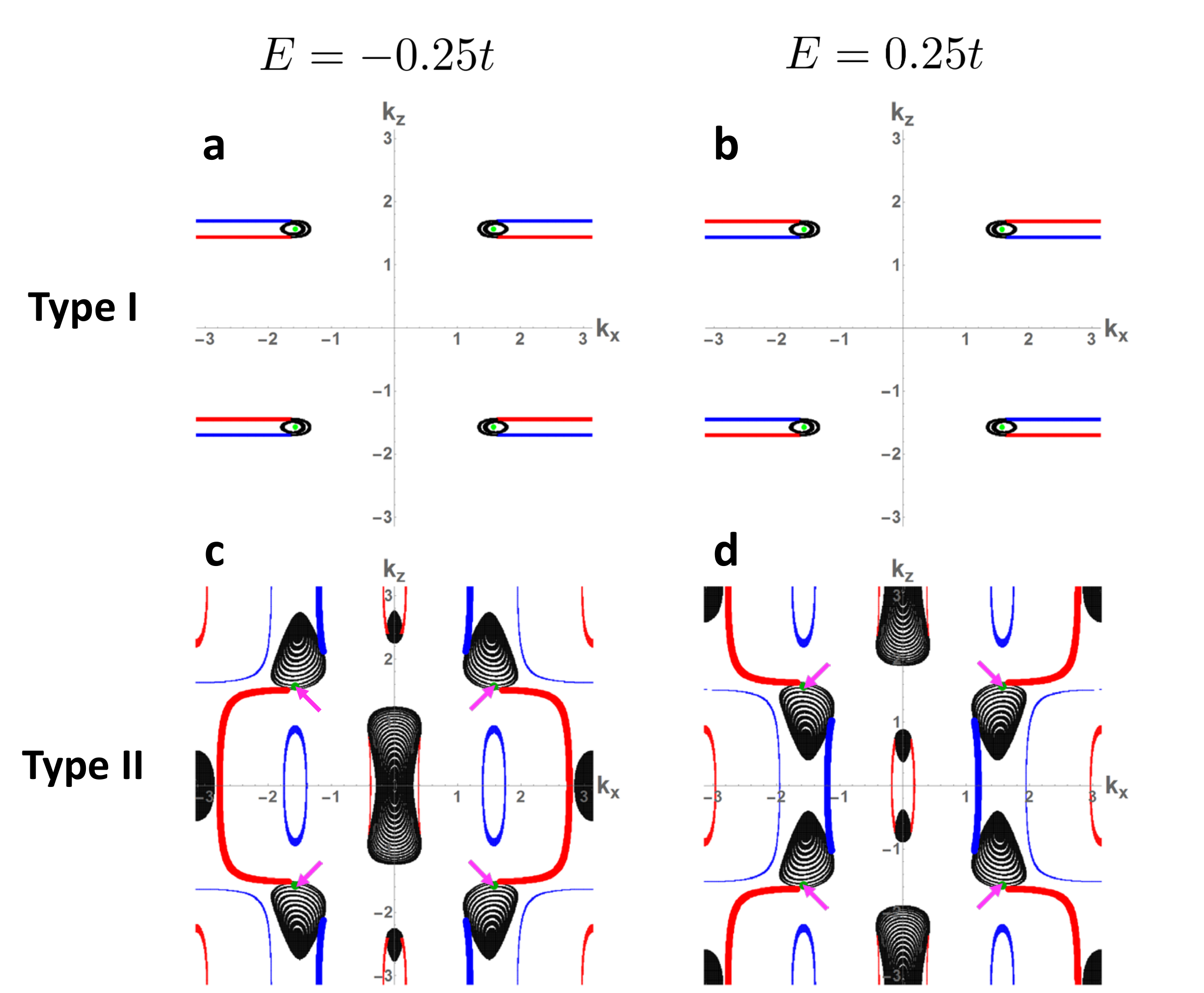}
	\caption{ {\bf{Fermi surface and Fermi arc configuration for type-I and type-II inversion-breaking Weyl semimetal.}}
	{\bf{a,b}} The Fermi surface and Fermi arc configuration for the Hamiltonian given in Eqn. (\ref{hinvkernel}) in the type-I limit ($\gamma = 0$) calculated in a slab geometry with $L = 50$ layers and with bulk parameters the same as in Fig. \ref{invBulk}a,d,g.  We show this calculation at constant energies: $E = -0.25t$ ({\bf{a}}) and $E = 0.25t$ ({\bf{b}}).  Here we see that Weyl nodes located at $(k_x,k_z) = (\pm \pi/2,\pm \pi/2)$ are connected by surface states (red and blue lines) to one of opposite chirality across the Brillouin zone in the $k_x$-direction.
	{\bf{c,d}} 
	The Fermi surface and Fermi arc configuration for the Hamiltonian given in Eqn. (\ref{hinvkernel}) in the type II limit ($\gamma = 2.4 t$) calculated in a slab geometry with $L = 50$ layers and with bulk parameters the same as in Fig. \ref{invBulk}c,f,i.  We show these for the same constant energies as above: ({\bf{c}}) and $E = 0.25t$ ({\bf{d}}).  The locations of the Weyl nodes are marked with pink arrows. We term the exponentially localized surface states that form closed loops ``track states".  Fermi arcs are shown as bold lines and connect Weyl nodes in the $k_z$-direction.
	}
	\label{inv_slab}
\end{figure*}

In order to study the Fermi arcs, we again construct a slab geometry by transforming the terms dependent on $k_y$ in Eqn. (\ref{hinvkernel}) into real space and considering a system with $L$ layers in the $y$-direction and infinite in the $x$- and $z$-directions.  In the type-I limit with $\gamma = 0$ shown in Fig. \ref{inv_slab}a and b, we find that away from $E = 0$, the projections of the nodes are enclosed by isolated small Fermi pockets. These pockets are connected to one another by topological Fermi arcs in the $k_x$-direction.  At $E = 0$, the top and bottom arcs are degenerate along the lines $k_z = \pm \pi/2$.  In a sense, this type-I ($\gamma = 0$) limit in the inversion-breaking model is effectively composed of two copies of a time-reversal-breaking Weyl semimetal separated by $\pi$ reciprocal lattice vectors along the $k_z$ direction.

\begin{figure*}
	\centering	\includegraphics[width=\textwidth]{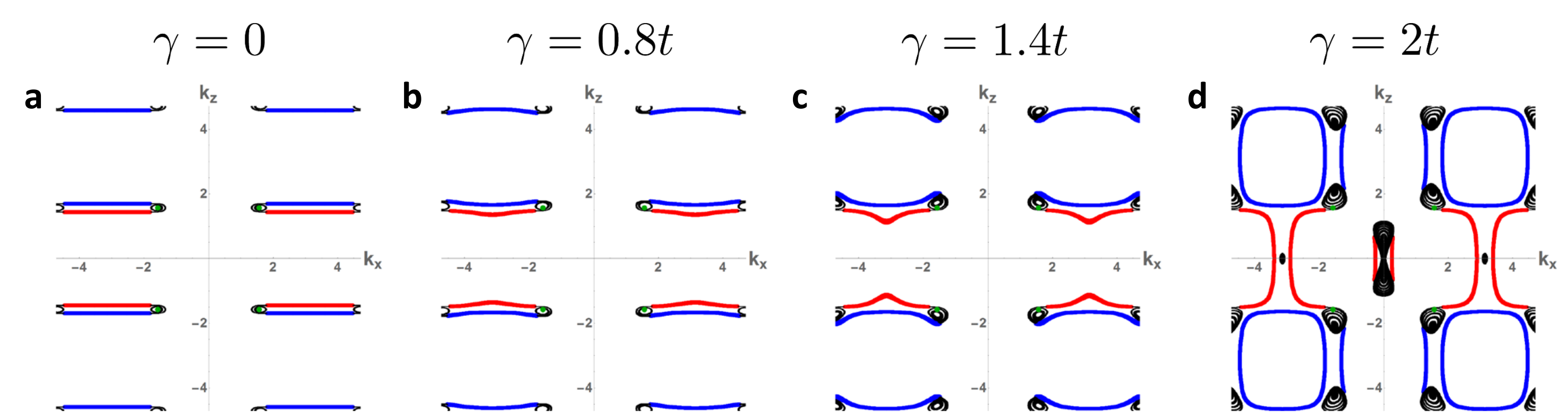}
	\caption{{\bf{Evolution of Fermi surface and Fermi arc configuration for inversion-breaking Weyl semimetal as a function of $\gamma$. }}
	{\bf{a-d}} The evolution of the Fermi surface and Fermi arc configuration in a slab geometry for Eqn. (\ref{hinvkernel}) with $k_0 = \pi/2$, $t_{x} = t/2$, $m = 2t$ calculated at constant energy $E = -0.25t$ for $\gamma = 0$ ({\bf{a}}), $\gamma = 0.8t$ ({\bf{b}}), $\gamma = 1.4t$ ({\bf{c}}), and $\gamma = 2.$ ({\bf{d}}) shown in an extended Brillouin zone where both $k_x$ and $k_z$ range from $-1.5 \pi$ to $1.5\pi$.  We see that at the critical point between a type-I and type-II ({\bf{d}}), the Fermi arcs that previously connected Fermi pockets in the $k_x$-direction now connect Fermi pockets in the $k_z$-direction and track states have formed on the bottom surface (blue) around the $(k_x,k_z) = (\pi,\pi)$ point.  
	}
	\label{extBzGamEv}
\end{figure*}

When $\gamma$ is increased to the type-II limit, the Fermi arc and bulk Fermi surface configuration in the inversion-breaking case is very different from the time-reversal-breaking model as we see in Fig. \ref{inv_slab}c and d.  The projections of the Weyl nodes are now enclosed by extended hole pockets for $E < 0$ (Fig. \ref{inv_slab}c) and electron pockets for $E > 0$ (Fig. \ref{inv_slab}d).  These Fermi pockets are connected by topological Fermi arcs, shown by thick red and blue lines, to pockets containing Weyl nodes of opposite chirality. Unlike in the type-I limit, here the Fermi arcs connect pockets along the $k_z$-direction rather than the $k_x$-direction.  One might expect that the transition point where the Fermi arcs connect nodes in one direction rather than another is concurrent with the transition point between a type-I and type-II Weyl semimetal and indeed our numerical calculations show that is the case (see Fig. \ref{extBzGamEv}).  Hence we see that for the same model with all other parameters held constant, merely tilting the nodes can lead to a dramatic recombination of the Fermi arcs and a qualitatively different pocket connectivity.

In Fig. \ref{inv_slab}c and d, we see that there are many states that are exponentially localized on the surface, however many of them form closed loops.  We term these closed loops ``track states"; they are degenerate in energy with the Fermi arcs but do not share their topology. Unlike Fermi arcs, track states form closed rather than open contours of surface states.
By investigating the evolution of the Fermi arc and Fermi surface configuration as a function of $\gamma$ (Fig. \ref{extBzGamEv}), we see that when the Fermi arc connectivity changes from the the $k_x$-direction to the $k_z$-direction, they leave behind track states around the $(k_x,k_z) = (\pi,\pi)$ point.

\section{Surface States: Topological and Track}
\begin{figure*}
	\centering	\includegraphics[width=\textwidth]{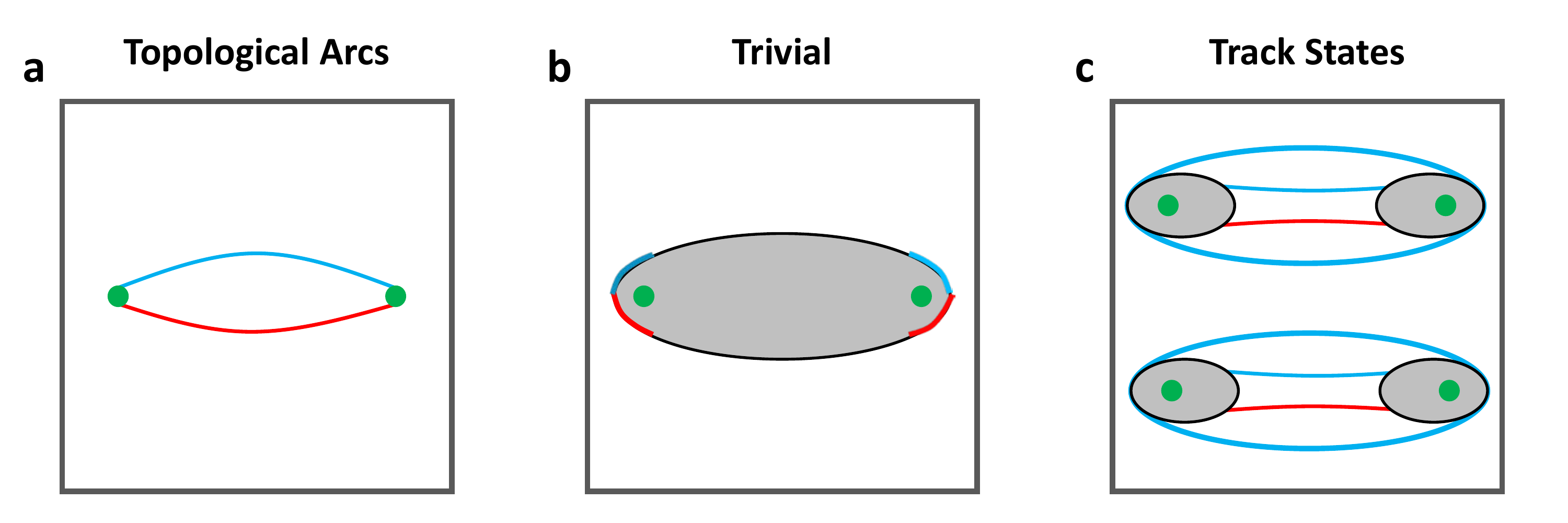}
	\caption{{\bf{Sketch of the three types of surface states in a topological Weyl semimetal. }}
	{\bf{a}} Two type-I Weyl nodes of opposite chirality connected by a Fermi arc on the top (red) and bottom (blue) surfaces.  In an arbitrary type-II TWS at an energy away from the Weyl energy, these arcs would connect Fermi pockets instead of nodes.
	{\bf{b}} A single Fermi pocket enclosing two nodes of opposite chirality.  Since no Gaussian surface can be constructed in a region that is both gapped and encloses only one node, the only possible surface states are trivial ones, shown in red and blue at the boundary of the pocket that hybridize with bulk states due to lack of topological protection. 
	{\bf{c}}  Pairs of Weyl nodes, two of each chirality with each node surrounded by a Fermi pocket.  The pockets are connected by Fermi arcs (thinner red and blue contours) as well as track states (thicker blue lines) on the bottom surface.  Note that states on opposite sides of a given loop of track states will disperse in opposite directions and so a Gaussian surface enclosing a given Fermi pocket will still have one net surface state of each chirality.
	}
	\label{topTrivTrack}
\end{figure*}

We briefly recapitulate the well-known topological argument for the existence of Fermi arcs in TWS.
Two 2D subspaces of the Brillouin zone can be treated like Gaussian surfaces enclosing a total charge of Berry curvature equal to the net chirality $\chi$ of the Weyl nodes inside.  It is a simple argument\cite{wanTurnVish} to show that these two 2D subspaces, often taken to be planes, must have Chern numbers \cite{PhysRevLett.49.405} that differ by $\chi$ enclosed.  This implies that one of these 2D planes in the BZ possesses $|\chi|$ more chiral edge modes on its boundary than the other plane. As we consider various families of such 2D planes in the Brillouin zone, these chiral edge modes trace out the open contours of surface states known as Fermi arcs that must terminate on Weyl nodes. In this way, there is a correspondence between the Berry curvature of the Weyl nodes, a bulk topological quantity, and these surface Fermi arcs (see sketch in Fig. \ref{topTrivTrack}a) that are also topological in nature.  

This argument is easily generalized\cite{Soluyanov2015} in a type-II Weyl semimetal by ensuring that the Gaussian surfaces one constructs in the Brillouin zone exist in regions that are gapped and therefore in general enclose Fermi pockets rather than bare nodes.  Hence in a type-II Weyl semimetal, Fermi pockets enclosing Weyl nodes will be connected by Fermi arcs.  However, in a type-II TWS, Fermi pockets can be extended enough to enclose multiple Weyl nodes.  If the net chirality enclosed by the Fermi pocket is zero, there will be no topologically protected Fermi arcs, as shown in the sketch in Fig. \ref{topTrivTrack}b.  The only surface states that can exist around this pocket will be trivial ones.

In the transition to a type-II TWS, a new surface state can appear, which we term a ``track state."   
These track states are degenerate with the Fermi arcs but do not share the topological properties of the arcs; they are generated as the connectivity of Weyl nodes changes as we tune the parameters of a system with multiple pairs of Weyl nodes, as shown in Fig. \ref{extBzGamEv} and discussed in the section on inversion breaking type-II Weyl semimetals above. They can exist in isolation in the Brillouin zone or can take part in the connectivity of Fermi pockets enclosing Weyl nodes, as illustrated in the sketch in Fig. \ref{topTrivTrack}c.   Lastly, we note that these track states can appear very similar to Fermi arcs when track states and arcs lie close together.  Caution must therefore be taken when analyzing the surface Fermi state configurations of type-II Weyl semimetals in DFT calculations or in ARPES data.

\section{Conclusions} We expect our summary of minimal models for type-I and type-II Weyl semimetals for both time-reversal-breaking and inversion-breaking cases will set the stage for future investigations of their properties in applied electric and magnetic fields. These models also provide a foundation for additional effects of repulsive and attractive interactions. Experimental discoveries of magnetism and superconductivity in Weyl semimetals could provide impetus for such theoretical studies.

\medskip

\section*{Acknowledgments} The authors would like to thank M. Kargarian and Y.-M. Lu for useful discussions.  We would also like to thank A. Kaminski for our collaboration with him on the ARPES study of type II TWS in MoTe$_2$, which inspired much of this work.  T. M. M. acknowledges funding from NSF-DMR-1309461 and N. T. was supported by the Center for Emergent Materials, an NSF MRSEC, under grant DMR-1420451

\bibliography{latticeModelRef}

\end{document}